\documentclass[12pt]{article}
\usepackage{amsmath,amssymb,epsfig,graphicx}
\makeatletter \@addtoreset{equation}{section} \makeatother
\renewcommand{\theequation}{\thesection.\arabic{equation}}
\newcommand{\ba}{\begin{array}}
\newcommand{\ea}{\end{array}}
\newcommand{\beq}{\begin{equation}}
\newcommand{\eeq}{\end{equation}}
\newcommand{\bea}{\begin{eqnarray}}
\newcommand{\eea}{\end{eqnarray}}

\def\bce{\begin{center}}
\def\ece{\end{center}}

\def\nonu{\nonumber}

\def\be{\beta}

\def\la{\lambda}

\newcommand{\tr}{\mbox{Tr}}

\def\eps6{{\displaystyle \mathop{\epsilon}^{6}}{}}

\def\nab6{{\displaystyle \mathop{\nabla}^{6}}{}}

\def\0{{\sst{(0)}}}
\def\1{{\sst{(1)}}}
\def\2{{\sst{(2)}}}
\def\3{{\sst{(3)}}}
\def\4{{\sst{(4)}}}
\def\5{{\sst{(5)}}}
\def\6{{\sst{(6)}}}
\def\7{{\sst{(7)}}}
\def\8{{\sst{(8)}}}

\def\ba{\begin{array}}
\def\ea{\end{array}}
\def\beq{\begin{equation}}
\def\eeq{\end{equation}}
\def\be{\begin{equation}}
\def\ee{\end{equation}}

\def\tr{\mathop{\rm tr}}

\def\la{\lambda}
\def\eps{\epsilon}

\def\ba{\begin{array}}
\def\ea{\end{array}}
\def\beq{\begin{equation}}
\def\eeq{\end{equation}}
\def\be{\begin{equation}}
\def\ee{\end{equation}}

\def\tr{\mathop{\rm tr}}

\def\la{\lambda}
\def\eps{\epsilon}

\newcommand{\bean}{\begin{eqnarray*}}
\newcommand{\eean}{\end{eqnarray*}}

\begin{document}
\thispagestyle{empty} \addtocounter{page}{-1}
\begin{flushright}
KIAS-P06039 \\
{\tt hep-th/0608160}\\
\end{flushright}

\vspace*{1.3cm}

\centerline{ \Large \bf Brane Configurations for}
\vspace{.3cm} 
\centerline{ \Large \bf Nonsupersymmetric Meta-Stable Vacua}
\vspace{.3cm} 
\centerline{ \Large \bf  in SQCD with Adjoint Matter} 
\vspace*{1.5cm}
\centerline{{\bf Changhyun Ahn}} 
\vspace*{1.0cm} 
\centerline{\it 
Department of Physics, Kyungpook National University, Taegu
702-701, Korea} 
\vspace*{0.8cm} 
\centerline{\tt ahn@knu.ac.kr} 
\vskip2cm

\centerline{\bf Abstract}
\vspace*{0.5cm}

We present the configurations of intersecting branes
in type IIA string theory corresponding to 
the meta-stable supersymmetry breaking 
vacua(hep-th/0608063) in the four-dimensional 
${\cal N}=1$ supersymmetric Yang-Mills theory coupled massive 
flavors with adjoint matter where the superpotential has three deformed
terms. 

\baselineskip=18pt
\newpage
\renewcommand{\theequation}
{\arabic{section}\mbox{.}\arabic{equation}}

\section{Introduction}

The dynamical supersymmetry breaking occurs at local long-lived 
meta-stable minima 
separated from supersymmetric vacua by potential barrier \cite{ISS}.
Those nonsupersymmetric meta-stable vacua exist in ${\cal N}=1$ SQCD
with massive fundamental matter(with masses much smaller than the dynamical
scale of the gauge sector).
The supersymmetry is 
broken by the rank condition(not all the F-term conditions
can be satisfied). Typically classical flat directions exist and can be 
lifted by quantum corrections which generate positive mass terms for the 
pseudomoduli leading to long-lived meta-stable vacua.   

A simple generalization of \cite{ISS} doesn't  always guarantee the existence
of local meta-stable minima at one-loop \cite{FU}. Somehow the direction 
which was not properly lifted by the one-loop potential should be removed.
In \cite{FU}, this can be accomplished by adding a set of singlets in the 
quiver gauge theory on fractional branes at the simplest example of an 
obstructed geometry. 
Other possible embedding is given by brane probes wrapping cycles of
local Calabi-Yau and corresponding ${\cal N}=1$ supersymmetric quiver 
gauge theories have a landscape of inequivalent meta-stable vacua and 
all the moduli are stabilized \cite{OO}. Although these theories have two
gauge singlet fields, one of the F-term conditions enables us to remove 
one of the degrees of freedom due to the coupling of bi-fundamentals and 
adjoint field of second gauge group in the superpotential. 

Geometrical approach for these meta-stable vacua using the brane 
configuration in type IIA string theory 
have many interesting aspects \cite{OO1,FGU,BGHSS}. 
For the example of \cite{ISS}, turning on the quark masses in the 
electric theory corresponds to deform the superpotential by adding 
a term in linear in a singlet field in the dual magnetic theory. 
In the brane configuration, 
this is equivalent to move D6-branes(with worldvolume 0123789 directions) 
in particular
directions. The misalignment of D4-branes(with worldvolume 01236 directions) 
connecting NS5'-brane(with worldvolume 012389 directions) can be 
interpreted as a nontrivial F-term conditions in the field theory side.
Some of these D4-branes can join and recombine 
the same number of D4-branes connecting 
NS5-brane(with worldvolume 012345 directions) 
which is equivalent to give nonzero vacuum expectation values of
dual quarks. The remaining D4-branes connecting NS5'-brane can move 
in other two directions freely and this is exactly the classical moduli 
space of nonsupersymmetric vacua of the field theory. 

For the application of \cite{ISS}, it is natural to ask what happens 
if we add an adjoint field into ${\cal N}=1$ SQCD with fundamental matter.
It turns out that  a local meta-stable minimum 
at one-loop doesn't exist without
adding any deformations.
In \cite{AGM}, by deforming the electric description of this theory
with an addition of gauge singlet terms, they found a meta-stable 
nonsupersymmetric long-lived vacuum.
It was crucial to consider the extra large massive adjoint field as well as
the mass term for quarks and the coupling term 
between an adjoint and quarks.   

In this paper,  
after the review of standard supersymmetric brane configuration
in electric and magnetic theories for ${\cal N}=1$ supersymmetric 
gauge theory with fundamental matter and an adjoint field, we would like to
construct the brane configuration  for nonsupersymmetric 
meta-stable vacua \cite{AGM}
by realizing the three deformation terms in the magnetic dual theory
geometrically. 

Some of the relevant works with \cite{ISS} in different context are present
\cite{BBO}-\cite{Forste}. For the standard description for the brane
construction with gauge theory, see the reference \cite{GK}.
We follow the notations of \cite{GK}.

\section{Brane configurations for the meta-stable vacua}

\subsection{Electric theory for supersymmetric brane configuration}

Before we are considering brane configuration for the
nonsupersymmetric meta-stable vacua in type IIA string theory, 
let us  
review the supersymmetric brane configuration for the 
${\cal N}=1$ $U(N_c)$ super Yang-Mills theory coupled 
$N_f$ massive chiral multiplets $Q, \widetilde{Q}$  
and a single adjoint field $X$ with a cubic superpotential 
$W_{ele}= \frac{1}{3} g_X \tr X^3$.

In the electric description, let us describe the 
configuration of $N_c$ D4-branes stretched 
between NS5-branes and an NS5'-brane along the 
$x^6$ direction, with $N_f$ D6-branes at values of $x^6$
that are between those corresponding to the positions of
the NS5-branes and NS5'-brane \cite{EGK,EGKRS,GK}. 
The power of a superpotential is related to the number of 
NS5-branes. For the cubic case, the number of NS5-branes is given by 2.

$\bullet$
Two NS5-branes  with worldvolume $(012345)$ living at a point 
in the $(6789)$ directions.

$\bullet$ 
One NS5'-brane with worldvolume $(012389)$ living at a point 
in the $(4567)$ directions.

$\bullet$
$N_f$ D6-branes with worldvolume $(0123789)$ living at a point 
in the $(456)$ directions.

$\bullet$
$N_c$ D4-branes with worldvolume $(01236)$ living at a point 
in the $(45789)$ directions.

The 4-6 strings describe $N_f$ chiral multiplets in the fundamental 
representation of $U(N_c)$ and the only massless mode arising from
4-4 strings  is the $U(N_c)$ worldvolume gauge field.
Here we define two complex coordinates
\bea
v \equiv x^4 + i x^5, \qquad w \equiv x^{8} + i x^9.
\nonu
\eea 

Initially, the two NS5-branes live at the same point in the $w$-plane.
Displacing them to 2 different points $w_1$ and $w_2$ provides
different possible configurations labelled by a set of non-negative 
integers $(r_1,r_2)$ where $r_1+r_2=N_c$. 
One can think of the locations $w_1$ and $w_2$ as determining
the vacuum expectation values of the adjoint field $X$
describing fluctuations of the D4-branes in the $w$-direction. 
There is a superpotential for the adjoint field $X$ of the form by resolving 
the cubic superpotential:
$W_{ele}=\frac{1}{3} g_X 
\tr X^3 + \frac{1}{2} m_X \tr X^2 +\la_X \tr X$. This superpotential
has 2 distinct minima in general. 
The integers $r_1$ and $r_2$ are the numbers of eigenvalues of the 
matrix $X$ residing in the different minima of the potential.
When $w_1$ and $w_2$ are distinct, $X$ is massive and the gauge group is 
broken $U(N_c) \rightarrow U(r_1) \times U(r_2)$.

The brane configuration for this theory is present in Figure 39 (a) of
\cite{GK}. Instead of drawing this in this paper, we will present 
the more important brane configuration for the magnetic theory 
in next subsection.  

\subsection{Magnetic  theory for supersymmetric brane configuration
\label{magnetic}}

The Seiberg dual, magnetic theory, 
can be obtained by moving NS5-brane to the other side of
NS5'-brane. For the present case, we do this procedure twice because
there exist two NS5-branes for our problem(coming from 
the cubic superpotential).
The brane configuration for the magnetic dual was identified in 
\cite{EGK,EGKRS,GK} and corresponding figure was presented in Figure 39 (b) 
of \cite{GK} or we redraw it in Figure 1 below.

$\bullet$ Two NS5-branes with worldvolume (012345)(denoted by red color) 

$\bullet$ One NS5'-brane with worldvolume (012389)(denoted by blue color)

$\bullet$ $N_f$ D6-branes with worldvolume (0123789)(denoted by dotted lines)

$\bullet$ $(2N_f-N_c)$ D4-branes between the 
NS5'-brane and NS5-branes: first kind of D4-branes

$\bullet$  $2N_f$ D4-branes between D6-branes and NS5'-brane: 
second kind of D4-branes.
Two D4-branes are connected to each of $N_f$ D6-branes \footnote{There is no
analog of the s-rule \cite{HW} 
for these D4-branes because two such D4-branes can be
separated in the $w$-direction \cite{EGK,EGKRS}.}.

The low energy theory on the D4-branes stretched between 
NS5'-brane and NS5-branes  is
a $U(\widetilde{N} =2N_f -N_c)$ gauge theory
with an adjoint superfield $Y$ coming from 4-4 strings
connecting D4-branes of the first kind, $N_f$ fundamental flavors
of quarks $q, \widetilde{q}$ coming from 
4-4 strings connecting D4-branes of the first kind to those of 
the second kind, and 2 magnetic meson fields $M_1$(which is equal to
$Q \widetilde{Q}$) and $M_2$(which is equal to $Q X \widetilde{Q}$)
each of which is an $N_f \times N_f$ matrix, coming from 
4-4 strings connecting D4-branes of the second kind. 
The extra meson degrees of freedom correspond to the motion of the 
$2N_f$ D4-branes in the $w$-direction along the D6-branes and 
NS5'-brane \footnote{In the gauge theory, this corresponds to putting 
$q=\widetilde{q} =Y=0$ and turning on arbitrary vacuum expectation 
values for the singlets $M_1$ and $M_2$ \cite{EGKRS}. The classical 
moduli space is $2N_f^2$ dimensional, not $4N_f^2$. 
One simply gets two copies of the 
moduli space of the magnetic theory for ${\cal N}=1$ SQCD with 
fundamentals \cite{ISS}. The explicit counting of $N_f^2$ 
massless modes was given in \cite{GK}: Figure 29 of \cite{GK}. }. 
The magnetic superpotential corresponding to 
the resolved electric superpotential \footnote{Although the electric 
superpotential has  extra adjoint $Z$-dependent terms \cite{EGKRS,GK}, 
if there exists 
a single NS5'-brane(i.e., quadratic superpotential in $Z$), then $Z$ is
massive and can be integrated out. This amounts to putting $Z=0$ in the 
superpotential. Therefore, there is no dual adjoint field in the 
magnetic superpotential. We will come to the finite massive case 
which will lead to quartic superpotential $(Q \widetilde{Q})^2$ 
in next subsection.} 
is given by \cite{Kutasov,KS,KSS}
\bea
W_{mag}  = 
\frac{1}{3} \widetilde{g}_Y \tr Y^3 + \frac{1}{2} \widetilde{m}_Y 
\tr Y^2 + \widetilde{\la}_Y \tr Y +
\tr \left( \frac{1}{2} \widetilde{m}_Y \widetilde{q} M_1 q + 
\widetilde{g}_Y \widetilde{q} M_2 q +
\widetilde{g}_Y M_1 q Y \widetilde{q}    \right).
\nonu
\eea
Also one can put the intermediate scale in the meson terms above 
by taking into account the mass dimension of meson and this scale 
relates the strong coupling scale of electric gauge group to  the 
strong scale coupling of its magnetic dual gauge group. 
The coefficient of beta function implies 
that $2\widetilde{N}-N_f$ is negative
when $N_f < \frac{2}{3} N_c$ and in the dual magnetic theory
we should have $\widetilde{N} > 0$. Therefore, for the region 
where the magnetic theory is IR free, 
\bea
\frac{1}{2} N_c < N_f < \frac{2}{3} N_c 
\nonu
\eea
the stable vacuum is present \cite{AGM} and we will study these vacua
in the brane configuration in next subsection.

For convenience, we present the supersymmetric 
brane configuration in magnetic theory in 
Figure 1 which will be our starting point.

\begin{figure}[ht]
   \epsfxsize=5in 
\centerline{\epsffile{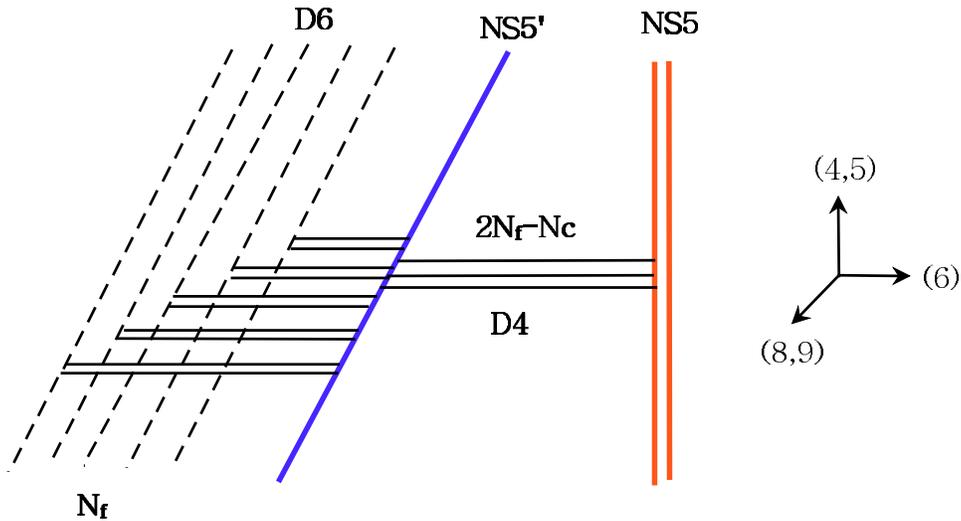}}
   \caption[FIG. \arabic{figure}.]{ 
The standard 
supersymmetric brane configuration for the magnetic 
$U(\widetilde{N}=2N_f-N_c)$ 
theory with $N_f$ flavors with an adjoint. 
There are also two gauge singlet fields.  This is  from a 
Figure 39 (b) of \cite{GK}. We emphasize here that the number of 
NS5-branes is two and the total D4-branes connecting to D6-branes is 
$2N_f$, not $N_f$. Two D4-branes are connected to each of 
$N_f$ D6-branes. }
\label{fig1}
\end{figure}

\subsection{Nonsupersymmetric brane configuration 
for the meta-stable vacua}

In \cite{AGM}, 
the dangerous extra flat directions which cannot be 
stabilized at one-loop
can be removed by deforming the original electric theory
where deformed mesonic terms are given by three terms
\bea
\Delta W_{ele} = \la_Q \tr Q X \widetilde{Q} + m_Q \tr Q \widetilde{Q} +
h \tr (Q \widetilde{Q})^2.
\nonu
\eea
The first term can be seen from ${\cal N}=2$ supersymmetric QCD where
when we give a vacuum expectation value to the adjoint field $X$, the 
hypermultiplets receive a mass. 
Rotating both NS5-branes relative to the D6-branes in ${\cal N}=2$ brane
configuration \cite{Barbon,GK} 
keeping the two NS5-branes and all the D6-branes parallel 
among themselves provides the change of the value for the coupling
$\la_Q$ between $X$ and quarks $Q$ and $\widetilde{Q}$. 
The second term is a mass deformation 
by turning on the mass matrix for the quarks that is an arbitrary
$N_f \times N_f$ complex matrix.
The last term can be obtained by integrating out an extra 
massive adjoint field $Z$
and as we take the limit $h \rightarrow 0$, we 
recover the usual ${\cal N}=1$ SQCD with an adjoint field $X$.
Rotating one NS5-brane by keeping other NS5-brane and all the D6-branes
fixed corresponds to the changing the mass of this extra adjoint chiral 
superfield $Z$. 

Let us describe the above deformations in the magnetic theory as follows:
1) Turning on the quark masses in the electric description 
corresponds to 
deforming the superpotential by adding a term in linear in $M_1$ since
the meson field $M_1$ is identified with 
$Q\widetilde{Q}$ in the electric description.
2) Changing the value of the Yukawa coupling between $X$ and the quarks
$Q$, $\widetilde{Q}$ 
corresponds to deforming the superpotential by adding 
a term in linear in $M_2$ because 
the meson field $M_2$ is identified with 
$Q X \widetilde{Q}$ in the electric description.
3) Finally, changing the mass of adjoint chiral superfield $Z$ 
corresponds to deforming the superpotential by adding 
a quadratic term in $M_1$.
Therefore, the full magnetic 
superpotential $W_{tot}$ 
with rescales can be written as  \cite{AGM} 
\bea
W_{mag}+ \Delta W_{mag} & = & 
\frac{1}{3} g_Y \tr Y^3 + \frac{1}{2} m_Y \tr Y^2 +\la_Y \tr Y +
\tr \left( h_1 \widetilde{q} M_1 q + h_2 \widetilde{q} M_2 q +
h_3 M_1 q Y \widetilde{q}    \right) \nonu \\
&-& 
h_1 m_1^2 \tr M_1 -h_2 m_2^2 \tr M_2 + m_3 \tr M_1^2. 
\label{superpotential}
\eea

Therefore, our main work is to find out the brane configuration 
by starting with the supersymmetric brane configuration in subsection  
\ref{magnetic} where the relevant superpotential is given by the first line
of (\ref{superpotential})
and adding the effects coming from the last three terms
in (\ref{superpotential}) and to see how it differs from the supersymmetric 
brane configuration.

One of the F-term conditions(a derivative of $W_{mag} + \Delta W_{mag}\equiv
W_{tot}$ 
with respect to
$M_2$), $F_{M_2}$, gives \cite{AGM}
\bea
q \widetilde{q} - m_2^2 {\bf 1}_{N_f} =0.
\label{Fterm}
\eea
Since the ranks of $q$ and $\widetilde{q}$ are at most 
$\widetilde{N}$, this equation cannot be satisfied if
the rank of the $N_f \times N_f$ matrix exceeds 
$\widetilde{N}$.
Turning on 
the value of the Yukawa coupling between $X$ and the quarks
$Q, \widetilde{Q}$ 
breaks the supersymmetry in the magnetic description.
This is the rank condition of \cite{ISS}.

The local minima of the tree-level $F$ and $D$-term potential 
from the field theory analysis
can be parametrized as \cite{AGM}
\bea
<q> &= & \left(
\begin{array}{c}
m_2 e^{\theta} {\bf 1}_{\widetilde{N}} \nonu \\
0
\end{array}
\right), \qquad 
<\widetilde{q}> =
 \left(
\begin{array}{cc}
m_2 e^{-\theta}  {\bf 1}_{\widetilde{N}}   &
0
\end{array}
\right), \qquad <Y>=0
\nonu \\
<M_1> &=& \frac{h_1}{2m_3}
 \left(
\begin{array}{cc}
(m_1^2-m_2^2)  {\bf 1}_{\widetilde{N}} & 0 \nonu \\
0 & m_1^2  {\bf 1}_{N_f-\widetilde{N}} 
\end{array}
\right), \nonu \\
<M_2>  &=&
 \left(
\begin{array}{cc}
- \frac{h_1^2}{2 h_2 m_3}(m_1^2-m_2^2)  {\bf 1}_{\widetilde{N}}  & 0 
 \\
0 & M_0  {\bf 1}_{N_f-\widetilde{N}} 
\end{array}
\right).
\label{solution}
\eea

Now we are ready to describe the brane configuration for the meta-stable
supersymmetry breaking vacuum characterized by (\ref{solution}).
Our starting point is the supersymmetric brane configuration in subsection 
\ref{magnetic} and describe brane dynamics for each deformation term in the 
superpotential (\ref{superpotential}). We will end up with the 
intermediate 
brane configuration  in Figure 2 after the following four steps.  

$\bullet$ Step 1. The $N_f$ D4-branes of second kind can move   
$w$-direction(since these D4-branes connecting 
D6-branes to the NS5'-brane are {\it not} rigid
and they can fluctuate in $w$-direction)  and arrive at $w=0$.

This is a reverse process when we are doing  Seiberg dual \cite{EGK,GK}. 
At this initial Step 1, we consider the vacuum structure where
the magnetic gauge group $U(\widetilde{N})$ is not broken by adjoint 
field $Y$. In brane language, remember that 
two nonnegative integers $r_1, r_2$ corresponding to 
the degeneracies of the eigenvalues of $Y$
where $r_1 + r_2 = \widetilde{N}$ and the locations $w_1$ and $w_2$ for
two NS5-branes in $w$-direction  corresponding to 
the eigenvalues for $Y$ determine the vacuum expectation value
for $Y$. When $r_1=0$, then we have $w_2=0$ implying that the 
$r_2=\widetilde{N}$ entries of $Y$ have zeros.
In other words, the two NS5-branes are located at $w=0$: {\it coincident } 
two NS5-branes (012345) at the origin $w=0$(denoted by two dotted red 
lines) in Figure 2.  
Other choices for the vacuum
expectation values of $Y$ don't change the tree-level
potential energy of the vacua \cite{AGM}.  

$\bullet$ Step 2. The NS5'-brane rotates in the $(v,w)$-plane freely since
all the D4-branes of first and second kinds
connecting to it are sitting at $w=0$. 

The {\it counterclockwise } 
rotation angle $\phi'$ when we 
look at from the right hand side of NS5'-brane is given by 
$\frac{1}{m_3(\phi')} = \tan \phi'$ \cite{Barbon,GK}.
This situation is in the intermediate status between NS5'-brane (012389)(the blue dotted line)
and tilted NS5'-brane(the blue solid line) of Figure 2. 
The relative angle $\phi'$ between the NS5'-brane and NS5-branes
encodes the mass of the adjoint chiral superfield 
$Z$ in the electric theory. In general, there is also a phase factor 
$e^{i \theta_{w}}$ in front of $\tan \phi'$ above
where $\theta_w$ is a rotation angle
in $w$-plane \cite{GK}.
In magnetic description, this rotation
induces the mass term for $M_1$.
Recall that one of the F-term conditions(a derivative of the 
superpotential $W_{tot}$ with respect to $M_1$), $F_{M_1}$,  
contains $m_3 M_1$ term.  

$\bullet$ Step 3. The $\widetilde{N}$ D4-branes move $v$-direction.

Among $N_f$ D4-branes of second kind, 
$\widetilde{N}$ D4-branes can combine with those D4-branes of first kind
and these combined D4-branes can leave NS5'-brane like as \cite{OO1,FGU}. 
In the electric 
brane description, the mass of quarks corresponds to the relative 
displacement between D6-branes and D4-branes in the 
$v$-direction(in the vertical direction of Figure 2).
The locations of D6-branes in the $v$-plane are the eigenvalues of
mass matrix. For equal masses \cite{AGM}, the $\widetilde{N}$ 
D4-branes are located at positive 
$v_{m_1}=-h_1 m_1^2$ which is a distance to the location for them 
from the origin in Figure 2. 
This contribution together with 
the contribution coming from  step 4 below will give us 
the final $\widetilde{N}$ entries for the meson $M_1$.
In magnetic description, this deformation
induces a linear term in $M_1$.
Note, in corresponding field theory side, 
that the expression of $-h_1 m_1^2 {\bf 1}_{N_f}$
is present in the F-term equation $F_{M_1}$.  
What happens for the remaining $(N_f-\widetilde{N})$ D4-branes of
second kind? The one end of these D4-branes is fixed at $(v,w)=(0,0)$ and the
other end is located at positive $v_{m_1}=-h_1 m_1^2$.
Remember that the D6-branes and NS5'-brane are { \it not }
parallel to each other and at an 
angle $\phi'$ at this stage. Therefore, those D4-branes connecting both
D6-branes and NS5'-brane
are rigid and cannot
move any directions freely.
Therefore, the $(N_f -\widetilde{N})$ entries of the 
vacuum expectation value of the meson $M_1$ correspond to 
the $v$ coordinate of D4-branes:$-\frac{v_{m_1}}{2m_3} {\bf 1}_{N_f -
\widetilde{N}}$. The overall coefficient $1/m_3$ is related to the rotation 
at Step 2.  

$\bullet$ Step 4. Both NS5 and NS5'-branes with same rotation angle
rotate in the $(v,w)$-plane.

The {\it counterclockwise} 
rotation angle $\phi$ when we 
look at from the right hand side of NS5-brane is given by 
$-h_2 m_2^2 = \cos \phi$ \cite{Barbon,GK}.
After this rotation, the final brane configuration 
is depicted in Figure 2.
In magnetic description, this 
induces a linear term in $M_2$.
In this process, $(N_f-\widetilde{N})$ D4-branes of second kind are not 
affected too much if $v_{m_1}$ is small 
because they cannot move unless the combined rotation of
$\phi'$ and $\phi$ give rise to the parallel between D6-branes and NS5'-brane. 
For general $v_{m_1}$, the $w$ coordinate of  end point 
of these D4-branes is not zero.
The NS5'-brane rotates in the $(v,w)$-plane with counterclockwise 
rotation angle $(\phi' + \phi)$ denoted by an angle between 
 NS5'-brane (012389)(the blue dotted line) and  tilted NS5'-brane(the
 blue solid line) 
in Figure 2
during the whole process.
In the field theory side, the F-term (\ref{Fterm}) can be partially
cancelled by a recombination of D4-branes in Step 3 above. In other words, 
this recombination corresponds to the fact that $\widetilde{N}$ entries
of the dual quarks $q, \widetilde{q}$ 
acquire nonzero vacuum expectation values  to minimize
the F-term.  
Therefore, the $\widetilde{N}$ entries of the 
vacuum expectation value of the meson $M_1$ coming from Step 3 above 
and Step 4 here
correspond to 
the $(v,w)$ coordinates of D4-branes:$-\frac{1}{2m_3} \left(v_{m_1}-
\frac{h_1}{h_2} 
\cos \phi \right) {\bf 1}_{
\widetilde{N}}$.  From the relation (\ref{Fterm}), 
one can write down 
$h_1 q \widetilde{q}$ arising from F-term equation $F_{M_1}$ as 
$\left(\frac{h_1}{h_2}\right) h_2 q 
\widetilde{q}=\left(\frac{h_1}{h_2} \right) 
h_2 m_2^2 {\bf 1}_{\widetilde{N}}$ and
the dimensions of $h_1 \cos\phi$ and $h_2 v_{m_1}$ are the same. 

Let us draw the intermediate brane configuration I in Figure 2 which shows the 
exact field theory results (\ref{solution}) except the information on the
vacuum expectation value for other gauge singlet $M_2$ that will be described
in Figure 3 and Figure 4.

\begin{figure}[ht]
   \epsfxsize=5in 
\centerline{\epsffile{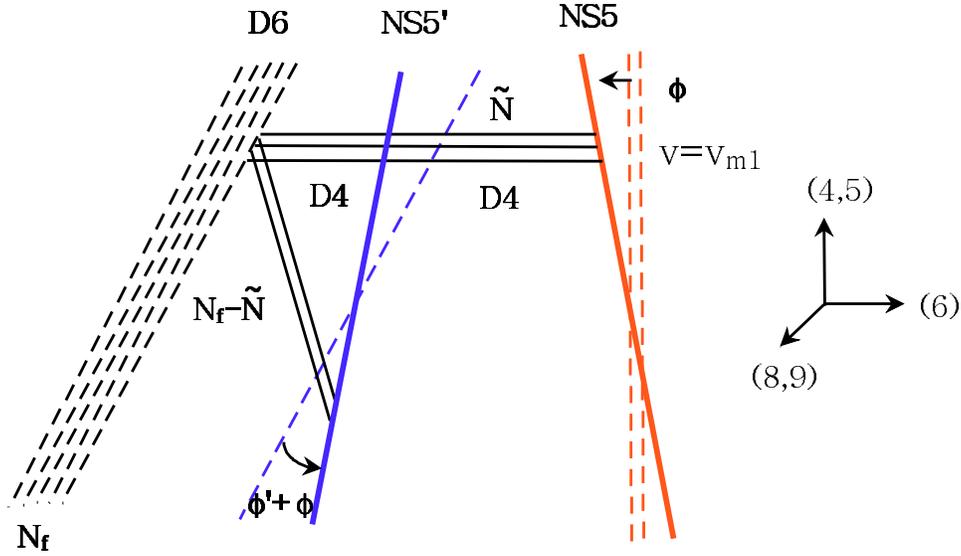}}
   \caption[FIG. \arabic{figure}.]{ 
The intermediate brane configuration I.
This brane configuration consists of tilted NS5'-brane(solid blue), 
tilted NS5-brane(solid red), D6-branes (0123789)(dotted black), and 
D4-branes connecting between D6-branes and NS5-brane or connecting
between D6-branes and NS5'-brane(solid black).
The three deformations are characterized by 
a rotation angle $\phi'$, a distance in $v$-direction $v_{m_1}$ and a
rotation angle $\phi$.
The vacuum expectation value for $M_1$ is fixed completely through  
this brane configuration. }
\label{fig2}
\end{figure}

So far, we have discussed about the moduli space of the magnetic theory 
coming from half of the $2N_f$ D4-branes of the second kind in the brane 
configuration. Now we describe the moduli space of the magnetic theory
coming from the remaining half of the $2N_f$ D4-branes of the second kind.
The process for this is as follows: 
The Step 1 and Step 3 are the same as before. The new feature appears 
in Step 2 and Step 4 denoted by Step 2' and Step 4' respectively below.

$\bullet$ Step 1. The {\it other} $N_f$ D4-branes of second kind can move   
$w$-direction and arrive at $w=0$.

$\bullet$ Step 2'. The NS5'-brane rotates in the $(v,w)$-plane freely since
all the D4-branes of first and second kinds
are sitting at $w=0$.

The {\it clockwise } 
rotation angle which is opposite to the previous rotation angle in Step 2
is denoted by $\phi''$ when we 
look at from the right hand side of NS5'-brane. This can be seen 
from the rotation from NS5'-brane (012389)(solid blue line) to tilted
NS5-brane(dotted blue line) in Figure 3. 
This rotation angle $\phi''$ will be fixed at the final stage.

$\bullet$ Step 3. The $\widetilde{N}$ D4-branes move $v$-direction.

For equal masses, the $\widetilde{N}$ 
D4-branes are located at positive 
$v_{m_1}=-h_1 m_1^2$ as before.

$\bullet$ Step 4'. Both NS5 and NS5'-branes rotate in the $(v,w)$-plane.
 
The {\it counterclockwise } 
rotation angle $\phi$ when we 
look at from the right hand side of NS5-brane is given by 
$-h_2 m_2^2 = \cos \phi = \cos \phi''$. In other words, 
the rotation of NS5'-brane st Step 2' is not independent but
depends on the rotation of NS5-branes.
Then the location of NS5'-brane doesn't change by the 
rotation from tilted NS5-brane(dotted blue line) to NS5'-brane(solid blue line) in Figure 3.
In this process, $(N_f-\widetilde{N})$ D4-branes of second kind
can move because
NS5'-brane is oriented into $w$ direction(the location of NS5'-brane does not
change during four steps) and therefore these D4-branes are
not rigid due to the parallel between D6-branes and NS5'-brane. 
In field theory analysis, the two equations of motion on $F_{q}$ and 
$F_{\widetilde{q}}$ do not fix the vacuum expectation value of $M_2$
completely. 
Therefore, the $(N_f-\widetilde{N})$ entries of the 
vacuum expectation value of the meson $M_2$ are not fixed and are parametrized
by pseudomoduli encoded in an arbitrary $(N_f-\widetilde{N}) \times 
(N_f-\widetilde{N})$ mesonic matrix $M_0$.
Moreover, 
the $\widetilde{N}$ entries of the 
vacuum expectation value of the meson $M_2$ correspond to 
the $(v,w)$ coordinates of D4-branes:$\frac{h_1}{2h_2 m_3} 
\left(v_{m_1}-\frac{h_1}{h_2} 
\cos \phi \right) {\bf 1}_{
\widetilde{N}}$ which is equal to minus $\widetilde{N}$ entries of 
the vacuum expectation value of $M_1$ multiplied by $h_1/h_2$.   
Note that the overall coefficient for this expression is 
different from the one in previous brane configuration where the 
corresponding rotation angle was $\phi'$.
According to the field theory analysis \cite{AGM}, 
the vacuum expectation value of $M_0$ in the quantum minimum
is nonzero \footnote{One of the pseudomoduli 
$\theta$ in the vacuum expectation
values $q$ and $\widetilde{q}$ is not manifest in the 
geometry we have considered.}.

We end up with the 
intermediate brane configuration II  in Figure 3 after above four steps.  

\begin{figure}[ht]
   \epsfxsize=5in 
\centerline{\epsffile{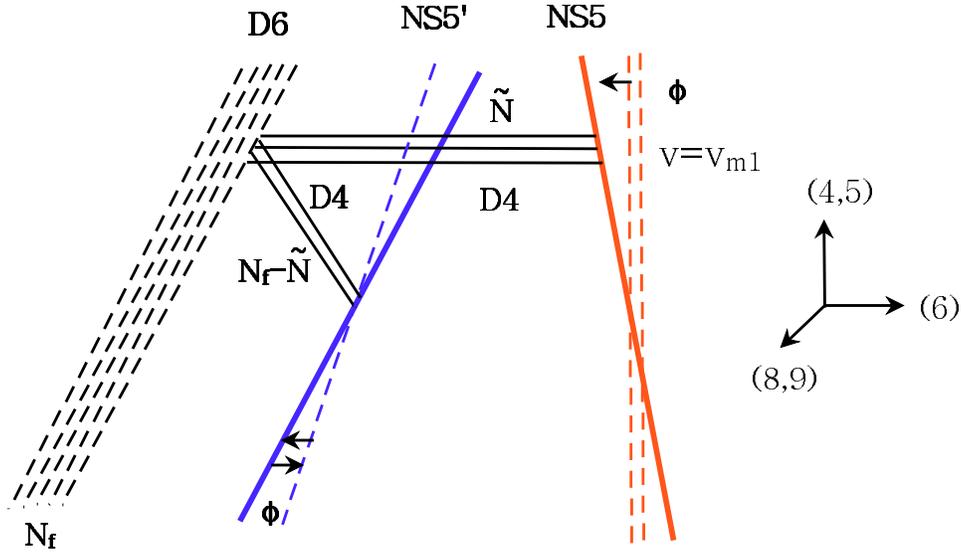}}
   \caption[FIG. \arabic{figure}.]{ 
The intermediate brane configuration II.
This brane configuration consists of NS5'-brane(solid blue), 
tilted NS5-brane(solid red), D6-branes (0123789)(dotted black), 
and D4-branes  connecting between D6-branes and NS5-brane or connecting
between D6-branes and NS5'-brane(solid black).
The vacuum expectation value for $M_2$ is partially fixed through  
this brane configuration.}
\label{fig3}
\end{figure}

Now let us combine the two different brane configurations I and II or describe
them simultaneously for the meta-stable vacua. 
Let us begin with Step 1 in brane configurations I and 
II at the same time. 
Then we continue Step 2 of brane configuration I while keeping the 
$N_f$ D4-branes appearing at the stage of Step 1 in brane configuration II intact. Next 
we do the Step 3 of both brane configurations I and II(i.e., turning on the mass term). 
Then, we end up with 
Step 4 of brane configuration I while 
remaining the 
above $N_f$ D4-branes intact in this case also.
At this stage, the only Step 1 and 3 of brane 
configuration II are done; splitting of $(N_f-\widetilde{N})$ D4-branes and 
$\widetilde{N}$ D4-branes.  
So one can imagine 
that there exists a brane configuration of Figure 2 together 
with these ``extra'' D4-branes.
 
We want to execute the remaining process of brane configuration II on these 
D4-branes.
Now we rotate NS5'-brane with an clockwise angle 
$(\phi'+\phi)$ in Figure 2 
and put it in its original position of Figure 3 without touching any other 
branes. This procedure can be interpreted as the relative magnitude 
of vacuum expectation values $M_1$ and $M_2$, in the field theory side. 
Then we perform the Step 2'
of brane configuration II with tilted $(N_f-\widetilde{N})$ D4-branes 
connecting to NS5'-brane and 
$\widetilde{N}$ D4-branes connecting to NS5-brane of 
Figure 2.  
Then the final procedure of Step 4' of brane configuration II can be 
done similarly. Note that  
although the positions of D4-branes connecting to NS5'-brane 
in Figure 2 are changed during the 
process of Step 2' of brane configuration II, after 
Step 4' of brane configuration II is done, they 
are located at the previous positions. Effectively they don't change in
this process. 
The Step 1 and 3 of brane configuration II were done already in
previous paragraph above.

So we can draw the final brane configuration corresponding to 
meta-stable vacua in Figure 4 which is a combination of brane
configurations I and II.
From this brane configuration, the classical moduli space for two
singlet fields without any deformations with dimension $2N_f^2$ 
is distributed to 
the vacuum expectation values for $M_1$ which is fixed 
completely(through 
the brane configuration I) and $M_2$ which is fixed 
partially(through the brane configuration II). 

\begin{figure}[ht]
   \epsfxsize=5in 
\centerline{\epsffile{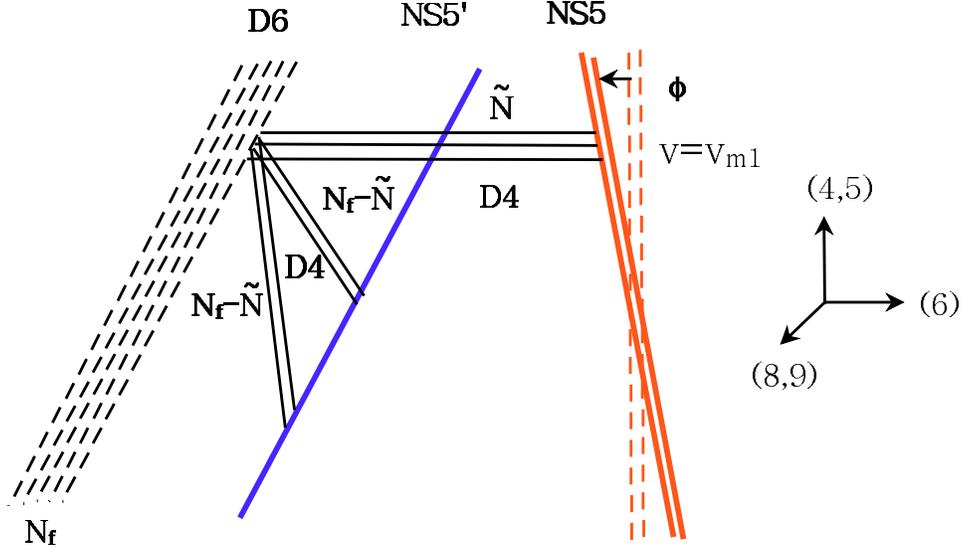}}
   \caption[FIG. \arabic{figure}.]{ 
The final brane configuration  corresponding to the field theory 
configuration (\ref{solution}).
This brane configuration consists of NS5'-brane (012389)(solid blue), 
tilted NS5-branes(solid red), D6-branes (0123789)(dotted black), 
and various D4-branes
 connecting between D6-branes and NS5-brane or connecting
between D6-branes and NS5'-brane(solid black). 
Half of D4-branes connecting to NS5'-brane 
can move in $w$ direction 
freely(corresponding to the vacuum expectation value $M_0$ of $M_2$
in (\ref{solution})). The other half of D4-branes connecting to NS5'-brane 
cannot move in $w$ direction and their $w$ coordinates intersecting NS5'-brane which are not
equal to zero depend on the rotation angles $\phi$ and $\phi'$. Their
$v$ coordinates intersecting D6-branes are specified by $v_{m_1}$. In
field theory side, this is equivalent to the fact that the vacuum
expectation value for $M_1$ is fixed completely. The
angle $\phi'$ is not seen explicitly in this figure.   The
three
deformation parameters are characterized by $\phi, \phi'$ and $v_{m_1}$. }
\label{fig4}
\end{figure}

As in the same spirit of \cite{OO1,FGU},
one can understand how the decay 
of the meta-stable vacuum into the 
supersymmetric vacua in this brane configuration. 
Let us move the upper $\widetilde{N}$ D4-branes toward to 
the NS5'-brane
such that there exist $\widetilde{N}$ D4-branes connecting NS5'-brane
and NS5-branes and $N_f$ D4-branes connecting the 
NS5'-branes and D6-branes in Figure 4. The final configuration for 
NS5-branes(dotted red lines) 
is  perpendicular to NS5'-brane(solid blue line). 
In this process, there exists an energy cost
which is equivalent to the height of potential barrier
between the nonsupersymmetric vacua and supersymmetric ones.
Now there exist $N_f$ D4-branes connecting the NS5'-brane and 
D6-branes and they can move in the $w$-direction.
The locations of these D4-branes in the $w$-direction
can be encoded by the vacuum expectation value of $M_2$.
The field theory result implies that 
there are supersymmetric configuration at \cite{AGM}
\bea
h_2 <M_2> ={m_2} 
\left( \frac{\widetilde{\Lambda}}{m_2}\right)^{\frac{N_f-2\widetilde{N}}{N_f-
\widetilde{N}}} \left( \frac{m_2}{m_Y} 
\right)^{\frac{\widetilde{N}}{N_f-\widetilde{N}}} {\bf 1}_{N_f}.
\nonu
\eea
This was obtained by taking the minmization of low energy effective 
superpotential which was obtained by integrating out 
the massive fields $M_1$ and $Y$ in (\ref{superpotential}) and using 
their equations of motion.
In the small $\frac{m_2}{\widetilde{\Lambda}}$ limit,
the vacuum expectation values above are much smaller than the scale 
$\widetilde{\Lambda}$ and much larger than mass scale $m_2$ of the 
magnetic theory.

\section{Comments}

Let us make some comments on the future directions.

$\bullet$ The magnetic configuration for the case where 
$k'$ NS5'-branes are connected by $\widetilde{N}(=N_f-N_c)$ D4-branes to
an NS5-brane on their right and by a single D4-brane to each of 
$N_f$ D6-branes on their left was given in \cite{EGKRS}.  
The low energy theory on the D4-branes of first kind 
has a gauge group $U(\widetilde{N})$ and an adjoint $Y'$ coming from 
4-4 strings connecting D4-branes of first kind, and $N_f$ fundamentals
$q$ and $\widetilde{q}$ coming from 4-4 strings connecting D4-branes of the 
first kind to those of the second kind. Finally, a single meson field
coming from 4-4 strings connecting D4-branes of the second kind
is coupled to the adjoint field and the quarks. 
For $k'=2$ case, the gauge theory analysis was given in \cite{ASY}.
It would be interesting to study whether there exist nonsupersymmetric 
meta-stable vacua by following the method of \cite{ISS}
and if not, how one can think of the possible 
deformations in the superpotential to make them stabilize 
at one-loop?

$\bullet$ As pointed out in \cite{FGU}, 
it might be possible to do 
a similar analysis for the nonsupersymmetric minima in orthogonal and 
symplectic gauge groups. 
The ${\cal N}=1$ SQCD with fundamentals and an adjoint field for these
gauge groups should have orientifold 4 plane, O4-plane with either 
plus D4-brane
charge or minus D4-brane charge. 
Similarly, the ${\cal N}=1$ SQCD with fundamentals 
and a symmetric(antisymmetric) tensor field for orthogonal(symplectic)
gauge group should have orientifold 6 plane.
We expect that for the former case with O4-plane, one should add
possible mesonic deformations in electric theory or magnetic theory because
the number of  magnetic meson fields is the same as the case of 
$U(\widetilde{N})$ gauge group. Also this is clear from the brane 
configuration we described in this paper.     

$\bullet$ Although there exist many field theory results \cite{ILS} 
where their Seiberg 
duals are known, it would be interesting to study meta-stable vacua for
some field theory 
results where D-brane descriptions are allowed. Some of the examples are 
presented in \cite{FGU} for unitary gauge groups.   Due to the presence of
many mesonic fields in the dual magnetic theory, a systematic analysis of the 
effect of deforming the theory by 
the addition of gauge invariant superpotential terms is needed. 

$\bullet$ As presented in \cite{Ooguri}, it might be possible to 
apply the method of \cite{ISS} to the more general quiver gauge theories.
The lesson from \cite{OO} implies that 
the coupling between dual quarks in the first gauge group factor and 
adjoint field of second gauge group factor gives rise to fix the direction of 
one of the mesonic fields. The immediate application in this direction is 
to take the product gauge group of orthogonal and symplectic gauge group with
appropriate matter contents and
this gauge theory appears in the brane configuration also. Or the 
orientifolded version of quiver gauge theory in \cite{OO} 
by taking the appropriate projections
correctly is possible.

$\bullet$ According to the observation of \cite{OV}, parallel to the standard
construction of \cite{EGK}, the geometry of ${\cal N}=1$ Seiberg duality
arises from wrapping D6-branes around 3-cycles of Calabi-Yau threefolds 
in type IIA string theory. In this process, the T-duality between 
the BPS states of ${\cal N}=2$ string theory  and ${\cal N}=1$ field theory
in four dimensions was crucial. So it is natural to ask how this geometric 
point of view in the meta-stable vacua brane configuration appears in various 
field theory results \cite{OV,Ahn}. In order to
do this, one needs to understand how T-duality works at the intersecting 
branes.
 
$\bullet$ There is also quantum effects in both electric 
and magnetic ${\cal N}=1$ SQCD theories. According to \cite{Witten},
the NS5-branes and NS5'-branes are deformed by the D4-branes 
ending on them. This bending is a quantum effect in the type IIA string 
description. For the problem we discussed in this paper, the careful analysis 
leads to the fact that 
only one combination of two $U(1)$ R-symmetries survives in the
quantum theory. The bending of the NS5-branes and NS5'-branes were given 
explicitly in \cite{EGKRS}.   
Although there exist some partial results in \cite{FGU}, the quantitative 
analysis remains open problem in the lifting the brane configuration of
M-theory. The recent paper \cite{BGHSS} might be helpful.

\vspace{.7cm}

\centerline{\bf Acknowledgments}

I would like to thank Freddy Cachazo, Sebastian Franco, Ryuichiro Kitano, 
David Kutasov, Kimyeong Lee, 
Yutaka Ookouchi, Chang Soon Park, Jaemo Park, 
Masaki Shigemori, Kyungsung Woo,
Ho-Ung Yee, 
and Piljin Yi for discussions.
In particular, I thank Yutaka Ookouchi for intensive discussions on 
ISS model \cite{ISS} and some comments on this paper.
This work was supported by grant No.
R01-2006-000-10965-0 from the Basic Research Program of the Korea
Science \& Engineering Foundation. 
I thank Hirosi Ooguri for my visit at 
Caltech Particle Theory Group  where
this work was undertaken and I thank KIAS(Korea Institute for 
Advanced Study) for hospitality.

\end{document}